\documentstyle[12pt]{article}

\begin{document}

\begin{center}

{\Large{\bf Coherent Effects under Suppressed Spontaneous Emission}\\ [5mm]
V.I. Yukalov} \\ [3mm]
{\it 
Bogolubov Laboratory of Theoretical Physics\\
Joint Institute for Nuclear Research, Dubna 141980, Russia \\
and \\
Instituto de Fisica de S\~ao Carlos, Universidade de S\~ao Paulo \\
Caixa Postal 369, S\~ao Carlos, S\~ao Paulo 135--970, Brazil}

\end{center}

\vskip 3cm

\begin{abstract}

An ensemble of resonance atoms is considered, which are doped into a medium
with well developed polariton effect, when in the spectrum of polariton
states there is a band gap. If an atom with a resonance frequency inside the
polariton gap is placed into the medium, the atomic spontaneous emission is
suppressed. However, a system of resonance atoms inside the polariton gap 
can radiate when their coherent interaction is sufficiently strong. Thus 
the suppression of spontaneous emission for a single atom can be overcome 
by a collective of atoms radiating coherently. Conditions when such 
collective effects can appear and their dynamics are analysed.

\end{abstract}

\vskip 2cm

{\bf PACS:} 42.50.Ct Quantum statistical description of interaction of 
light and matter -- 42.70.Qs Photonic bandgap materials

\newpage

\section{Introduction}

Inhibition of spontaneous emission has first been predicted [1,2] to occur 
in three-dimensional periodic structures, in which, due to periodicity, an
electromagnetic band gap develops. As a result, the spontaneous emission
with a frequency inside the band gap is rigorously forbidden. This type of
matter, where photon band gap arises because of the structural periodicity
in real space, has been called photonic band-gap materials [3].

Band gaps also appear in natural dense media due to photon interactions
with optical collective excitations, such  as phonons, magnons, excitons,
and so on [4,5]. This kind of a gap is termed the polariton band gap. If a
single atom is placed in a medium with the polariton band gap, and the
atomic resonance frequency lies inside the gap, then the atom spontaneous
emission can be suppressed [6,7].

Physical reasons for the effect of suppressed spontaneous emission 
are  rather similar for both types of materials, the difference being 
in the  nature of scatterers light interacts with. In artificial photonic 
band-gap materials, a suppression of the photon density of states over 
a  narrow frequency range results from multiple photon scattering by 
spatially correlated scatterers. In natural dense media, such as 
dielectrics or semiconductors, a frequency gap for electromagnetic 
modes  develops as a result of photon scattering by optical collective 
excitations. The suppression of spontaneous emission can be explained 
as follows. Let us imagine an excited atom in a medium, with the
atomic transition frequency within the prohibited gap. The atom tends to
become deexcited by emitting a photon. However, since the propagation of
photons inside the gap is prohibited, the emitted photon is scattered back
and is again absorbed by the atom. This means that the atom cannot become
deexcited or, in other words, that spontaneous emission is suppressed. 
Hence, this effect is characterized by the condition
$$
\lim_{t\rightarrow\infty}\; s(t) > \; -1
$$
for the average population difference $s\equiv<\sigma_i^z>$ that is a 
statistical average of the Pauli operator $\sigma_i^z$. In the described 
physical picture, the limit $t\rightarrow\infty$ is, of course, 
conditional simply implying that the suppression of spontaneous emission 
happens during times much longer than the emission time $\gamma^{-1}$, 
where $\gamma$ is the spontaneous emission rate. The complete suppression 
for infinite times is, certainly, never possible since there always exist 
spontaneous photons with frequencies out of the polariton band gap, so 
that sooner or later an atom becomes deexcited via nonresonant photons. 
However, for sufficiently large polariton gaps, such nonresonant 
relaxation processes are very slow. The characteristic relaxation time
of a nonresonant process occurring through a frequency gap $\Delta$ can 
be estimated [8,9] as $\gamma^{-1}(1+\Delta^2/\gamma^2)$. Accepting the 
values typical of the majority of semiconductors [4,5,10--12], for the
polariton gap one has $\Delta\sim 10^{13}$s$^{-1}$ while the emission 
rate is $\gamma\sim 10^8-10^9$s$^{-1}$. Hence the characteristic time of 
relaxation through nonresonant processes is $0.1-100$ s, which is 
essentially longer than typical emission times of free atoms. In this 
paper, we do not take into account the nonresonant deexcitation of atoms, 
which is justified for rather long times as estimated above.

Even though spontaneous emission of a singe atom in a medium is suppressed,
the situation for an ensemble of atoms can be different. When a collective
of resonance atoms is doped into the medium with a band gap, an impurity
band can be formed within the prohibited gap. This happens in photonic
band-gap materials [13] as well as in media with the polariton gap [14,15].
For the collection of identical resonance atoms with spacing much shorter
than the transition wavelength, electromagnetic effective interaction takes
place due to photon exchange. For closely spaced atoms with transition
frequencies in the gap, the virtual photon exchanges are of such energy that
the atoms do not experience the existence of the gap [16,17]. When an 
impurity band is formed, electromagnetic radiation becomes possible. 
Sufficiently strong effective interactions collectivize the resonance 
atoms that start radiating coherently. Thus, the suppression of 
spontaneous emission for a single atom can be overcome by a collective of 
atoms. Hence the suppressed light can be liberated, at least partially. 
But what would be the dynamics of such a collective radiation?

In the case of photonic band-gap materials, spontaneous emission near the
edge of a photonic band-gap has been considered [18,19] for a point-like
model with the radiation wavelength much larger than the model linear size,
$\lambda\gg L$. For polariton-gap media, the dynamics of collective liberation
has not yet been properly studied. Moreover, the consideration of 
point-like models, with $\lambda\gg L$, in the case of suppressed 
spontaneous emission cannot be accepted as realistic because of the 
follows. Assume that an excited atom, with a transition frequency within 
the prohibited photonic gap, is placed into a bandgap material. A photon 
emitted by the atom, before being reflected back, travels at the distance 
called the localization length $l_{loc}$ that is about several wavelengths, 
$l_{loc}>\lambda$. It is only when the localization length is much 
shorter than the sample size, $l_{loc}\ll L$, there is sense of talking 
that light can be localized, hence spontaneous emission be suppressed. 
While for a concentrated point-like system, with $\lambda\gg L$, one has 
even more $l_{loc}>\lambda\gg L$. That is, an emitted photon will safely
quit the sample, being never reflected back. Therefore the localization 
of light and suppressed emission principally cannot occur in a point-like 
system.

In the present paper, we consider a system of $N$ resonance atoms doped into
a polariton-gap medium, in which spontaneous emission of a single atom is
suppressed. The general case of arbitrary wavelengths is studied, which makes
it necessary to take into account local--field effects. The dynamics of
radiation is carefully analysed. Two basic points difference the present 
consideration from the works of other authors: First, a realistic 
situation is analysed, when the radiation wavelength can be (and {\it 
must be}, for the spontaneous emission be suppressed and light be 
localized) much shorter than the sample size. Second, the transition 
frequency is assumed to lie not at the edge of the gap but {\it deeply 
inside} the prohibited bandgap.
 The system of units is used with the Planck
constant $\hbar\equiv 1$.

\section{Evolution Equations}

Aiming at studying the dynamics of more or less realistic system, we start
with the Hamiltonian
\begin{equation}
\label{1}
\hat H = \hat H_a + \hat H_f + \hat H_{af} +\hat H_m +\hat H_{mf}
\end{equation}
containing the following parts. The Hamiltonian of two-level resonance atoms
\begin{equation}
\label{2}
\hat H_a =\frac{1}{2}\; \sum_{i=1}^N\; \omega_0(1 +\sigma_i^z) \; ,
\end{equation}
with the transition frequency $\omega_0$. The radiation-field Hamiltonian
\begin{equation}
\label{3}
\hat H_f =\;\frac{1}{8\pi}\; \int\; \left ( {\bf E}^2 + {\bf H}^2
\right )\; d{\bf r} \; ,
\end{equation}
with electric field ${\bf E}$ and magnetic field ${\bf H}={\bf\nabla}\times
{\bf A}$. The vector potential ${\bf A}$ is assumed to satisfy the Coulomb
gauge calibration ${\bf\nabla}\cdot{\bf A}=0$ and the commutation relation
$$
\left [ E^\alpha({\bf r}, t),\; A^\beta({\bf r}',t)\right ] =
4\pi ic\delta_{\alpha\beta}\delta({\bf r}-{\bf r}')\; ,
$$
where $c$ is the light velocity. The atom-field interaction is given by the
standard dipole Hamiltonian
\begin{equation}
\label{4}
\hat H_{af} = \; - \; \sum_{i=1}^N\; \left (
\frac{1}{c}\; {\bf j}_i\cdot {\bf A}_i + {\bf d}_i\cdot {\bf E}_{0i}\right 
)\; , 
\end{equation}
in which ${\bf A}_i\equiv{\bf A}({\bf r}_i,t)$; the transition-current and the
transition-dipole operators, respectively, are
\begin{equation}
\label{5}
{\bf j}_i= i\omega_0\left ({\bf d}\sigma_i^+ - {\bf d}^*\sigma_i^-\right )\; ,
\qquad {\bf d}_i ={\bf d}\sigma_i^+ +{\bf d}^*\sigma_i^- \;,
\end{equation}
where ${\bf d}$ is the transition dipole and $\sigma_i^\pm$ are the ladder
operators; and ${\bf E}_{0i}$ is an external field, if any. The Hamiltonian of
matter, $\hat H_m$, can be modelled by a collection of oscillators [6,7].
And matter-field interactions are given by the Hamiltonian
\begin{equation}
\label{6}
\hat H_{mf} =\; -\; \frac{1}{c}\; \sum_{i=1}^{N_0}\; {\bf j}_{mi}\cdot
{\bf A}_i\; ,
\end{equation}
where ${\bf j}_{mi}$ is a current produced by matter at the point ${\bf r}_i$;
$N_0$ being the number of oscillators modelling the matter.

When considering stationary properties of electromagnetic field, one usually
passes to the momentum-energy representation by expanding operators in
Fourier series. However this representation is not as convenient for studying
dynamical properties of a system of atoms, especially when their transition
wavelength is much shorter than the system size. Therefore we shall employ
here the space-time representation whose brief description can be found in
book [20] and detailed analysis in Refs. [21,22].

The Heisenberg equations of motion, with the Hamiltonian (1), yield for the
atomic variables the equations
$$
\frac{d\sigma_i^-}{dt} = - i\omega_0\sigma_i^- +\left ( k_0{\bf d}\cdot
{\bf A}_i - i{\bf d}\cdot{\bf E}_{0i}\right )\sigma_i^z\; ,
$$
\begin{equation}
\label{7}
\frac{d\sigma_i^z}{dt} = - 2\left ( k_0{\bf d}\cdot{\bf A}_i -
i{\bf d}\cdot{\bf E}_{0i}\right )\sigma_i^+ -
2\left ( k_0{\bf d}^*\cdot {\bf A}_i +
i{\bf d}^*\cdot{\bf E}_{0i}\right )\sigma_i^-\; ,
\end{equation}
where $k_0\equiv\omega_0/c$. The Heisenberg equations for the electric field
and vector potential lead to the Maxwell operator equations, combining which
with the Coulomb gauge calibration, one gets
\begin{equation}
\label{8}
\left ( {\bf\nabla}^2 -\; \frac{1}{c^2}\; \frac{\partial^2}{\partial t^2}
\right ){\bf A} = \; -\; \frac{4\pi}{c}\; {\bf J} \; ,
\end{equation}
with the density of current
\begin{equation}
\label{9}
{\bf J}({\bf r},t) = \sum_{i=1}^N\; {\bf j}_i(t)\delta({\bf r}-{\bf r}_i) +
\sum_{i=1}^{N_0}\; {\bf j}_{mi}(t)\delta({\bf r}-{\bf r}_i) \; .
\end{equation}
The formal operator solution to Eq. (8) reads
\begin{equation}
\label{10}
{\bf A}({\bf r},t) ={\bf A}_{vac}({\bf r},t) +\; \frac{1}{c}\; \int \;
{\bf J}\left ({\bf r}',t-\;\frac{1}{c}\; |{\bf r}-{\bf r}'|\right )\;
\frac{d{\bf r}'}{|{\bf r}-{\bf r}'|} \; ,
\end{equation}
where ${\bf A}_{vac}$ is a solution of the related uniform equation.
Substituting the density of current (9) into Eq. (10) results in the vector
potential
\begin{equation}
\label{11}
{\bf A} ={\bf A}_{vac} +{\bf A}_{rad} + {\bf A}_{mat}
\end{equation}
consisting of the potential ${\bf A}_{vac}$ of vacuum fluctuations, potential
${\bf A}_{rad}$ produced by radiating atoms, and the potential ${\bf A}_{mat}$
induced by the medium,
$$
{\bf A}_{rad}({\bf r}_i,t) = \sum_j^N\; \frac{1}{c\;r_{ij}}\;
{\bf j}_j\left ( t -\; \frac{r_{ij}}{c}\right ) \; ,
$$
\begin{equation}
\label{12}
{\bf A}_{mat}({\bf r}_i,t) = \sum_j^{N_0}\; \frac{1}{c\;r_{ij}}\;
{\bf j}_{mj}\left ( t -\; \frac{r_{ij}}{c}\right ) \; ,
\end{equation}
where the summation excludes the terms with $j=i$, and the notation
$r_{ij}\equiv|{\bf r}_{ij}|, \; {\bf r}_{ij}\equiv{\bf r}_i-{\bf r}_j$, is used.
It is worth emphasizing that the relation (10) is an exact expression 
immediately following from the evolution equation (8) and identical to 
the latter. No far-zone approximation is involved here. This 
approximation could be done [23] by expanding $|{\bf r}-{\bf r}'|^{-1}$
in the integrand of Eq. (10) in powers of $r'/r$, assuming that 
$r\equiv|{\bf r}|\gg r'\equiv |{\bf r}\; '|$. But we do not invoke here 
such an expansion. Substituting Eqs. (10) to (12) into the Hamiltonian 
(4) and the evolution equations (7), it is easy to notice that all these 
equations contain an effective dipole-dipole coupling of atoms 
responsible for near-zone local-field effects [24]. Expression (10) is 
not a final solution for the observable field but it is an {\it operator 
relation} having sense only in the combination with the Heisenberg 
equations (7). The system of Eqs. (7) and (10) follows directly from the 
initial Hamiltonian (1) and is often employed in optics [20].

Let us introduce the quantities
\begin{equation}
\label{13}
u_i(t)\equiv\; <\sigma_i^-(t)>\; , \qquad s_i(t)\equiv\; <\sigma_i^z(t)>\; ,
\end{equation}
in which the angle brackets imply the statistical averaging over the {\it
atomic degrees of freedom} only, not touching the vacuum and matter degrees
of freedom. For the double correlators, we shall employ the mean-field-type
decoupling
\begin{equation}
\label{14}
<\sigma_i^\alpha\sigma_j^\beta>\; =\; <\sigma_i^\alpha><\sigma_j^\beta>
\qquad (i\neq j) \; .
\end{equation}
It is worth emphasizing that, since only the atomic degrees of freedom have
been involved in the averaging, the decoupling (14) does not kill quantum
effects related to the vacuum and the matter degrees of freedom, which have
not yet been averaged out. The variables corresponding to these degrees of
freedom will enter the equations of motion for the quantities (13) as operator
variables or they can be modelled by stochastic variables [25]. This is why
the decoupling (14) may be called the {\it stochastic mean-field
approximation}.

Another problem concerns the way of treating the retardation appearing
in Eqs. (12). We cannot use the Markov approximation just ignoring this
retardation since this would eliminate all local-field effects that are so
important for realistic many-atom systems with wavelength shorter than the
system size [26]. But we may employ the Born approximation
\begin{equation}
\label{15}
<\sigma_j^-\left ( t -\; \frac{r_{ij}}{c}\right ) > \; =\;
u_j(t) \exp( ik_0r_{ij})
\end{equation}
retaining the account of local fields. Recall that considering resonance
atoms, one always keeps in mind that the intensities of interactions of
atoms with all fields are assumed to be much smaller than the transition
frequency $\omega_0$. If this were not so, then the notion of resonance
atoms as such would loose its sense, because strong fields, essentially
shifting the atomic energy levels, would completely change the
classification of these levels and the values of transition frequencies.
Under the assumption that the transition frequency $\omega_0$ is the
largest energy scale, from Eqs. (7) it immediately follows that the
approximation (15) is completely justified [20--22]. This can also be
called the quasirelativistic approximation since in the nonrelativistic
limit, when $c\rightarrow\infty$, Eq. (15) becomes an identity.

The equations for the quantities (13) follow from Eqs. (7) where we
introduce the notation for the effective field
\begin{equation}
\label{16}
f_i(t)\equiv k_0\; < {\bf d}\cdot{\bf A}_{rad}({\bf r}_i,t)> -
i{\bf d}\cdot{\bf E}_{0i} +\xi_i(t) \; ,
\end{equation}
in which the last term is the random field
\begin{equation}
\label{17}
\xi_i(t) \equiv k_0{\bf d}\cdot \left [ {\bf A}_{vac}({\bf r}_i,t) +
{\bf A}_{mat}({\bf r}_i,t)\right ] \; .
\end{equation}
As usual, we also need to take into account the level width $\gamma_1$ and
the line width $\gamma_2$. The introduction of these widths is done in 
the standard way [23,25,27] keeping in mind that the values of $\gamma_1$ 
and $\gamma_2$, in general, differ from the emission rate $\gamma$ of a 
free atom. This is because the longitudinal relaxation rate $\gamma_1$ is 
assumed to effectively include the influence of matter which atoms are 
dopped in, for instance, the influence of lattice vibrations. The 
transverse relaxation rate $\gamma_2$ differs from $\gamma$ because of 
the same reason of incorporating the influence of matter, for example, by
taking account of dynamic and, partially, of inhomogeneous broadening. Thus,
$\gamma_1$ and $\gamma_2$ are to be treated as phenomenological 
parameters introduced in the commonly accepted way [23,25,27]. In 
this way, we obtain the equations for the average transition amplitude
\begin{equation}
\label{18}
\frac{du_i}{dt} = \; - (i\omega_0 +\gamma_2) \; u_i + f_is_i \; ,
\end{equation}
and the average population difference
\begin{equation}
\label{19}
\frac{ds_i}{dt} =\; - 2( u_i^*f_i + f_i^* u_i) - \gamma_1(s_i - s_0) \; .
\end{equation}
The last term in Eq. (19) allows for the suppression of spontaneous
emission in the case of a single atom. Really, to return to the latter
case, we have to put $f_i=0$. Then we get the equation for a single atom,
with the solution $s_i(t)=s_0$, as is discussed in the Introduction. But,
in the case of an ensemble of atoms, the presence of nonlinear terms
related to the effective field (16) can make the solutions to the evolution
equations quite nontrivial. These equations are to be complimented by an
equation either for $u_i^*$ or for $|u_i|^2\equiv u_i^*u_i$. For the latter
equation, we find
\begin{equation}
\label{20}
\frac{d|u_i|^2}{dt} =\; - 2\gamma_2 |u_i|^2 + s_i
(u_i^*f_i + f_i^* u_i) \; .
\end{equation}
Equations (18) to (20) are the main evolution equations we shall consider.
These are the stochastic differential equations since they contain the
random field (17) including the vacuum and matter degrees of freedom.
Electromagnetic fluctuations coupled with collective excitations of a
medium are called polaritons [4,5]. Therefore Eq. (17) represents the
polariton filed acting on atoms.

Strictly speaking, one should distinguish two different physical cases, 
depending on the relation between the longitudinal relaxation time 
$T_1\equiv\gamma_1^{-1}$ and the characteristic time of nonresonant 
processes $T_n=\gamma^{-1}(1+\Delta^2/\gamma^2)$ discussed in the 
Introduction. As is explained in the latter, it makes sense of talking 
about the suppressed spontaneous emission of an atom only for times $t\ll 
T_n$. This suppression can be either dynamic, when $T_1\ll T_n$, or 
static, when $T_1\sim T_n$. In the first case, one has to retain the 
parameter $\gamma_1$ in the evolution equations, while in the second 
case, one has to set $\gamma_1=0$. For generality, we shall consider both 
these situations starting with the more general case of finite 
$\gamma_1$, from which it is easy to pass to the particular case of 
$\gamma_1=0$.

\section{Scale Separation}

In what follows, we consider the case when an external field is used only
for exciting atoms at the initial time, but after this the field is
switched off, ${\bf E}_{0i}=0$. It is worth saying a few words about the
possibility of exciting resonance atoms whose transition frequency lies
in the polariton band gap. There can be several ways of doing this. The
first way is by exciting the atoms with a short pulse of a resonance
external field. The amplitude of the latter is to be sufficiently large
since a weak electromagnetic wave with a frequency within the polariton
gap cannot propagate through the medium because of total reflection.
However, when the incident field is strong enough, a monocromatic wave
can penetrate into the medium with a polariton gap due to nonlinear
effects [28,29]. Another possibility for exciting the atoms in a polariton
gap could be through a third level with the related transition frequency
lying outside the gap. It could also be possible to get the population
inversion at the desired level through a set of levels outside the gap,
by using nonresonance pumping.

It is convenient to define the arithmetic averages
\begin{equation}
\label{21}
u(t) \equiv \frac{1}{N}\; \sum_{i=1}^N\; u_i(t) \; , \qquad
s(t) \equiv \frac{1}{N}\; \sum_{i=1}^N\; s_i(t)
\end{equation}
of the variables (13) and also the average polariton field
\begin{equation}
\label{22}
\xi(t) \equiv\frac{1}{N}\; \sum_{i=1}^N\; \xi_i(t) \; .
\end{equation}
Let us introduce the collective frequency $\Omega$ and the collective
width $\Gamma$, respectively,
\begin{equation}
\label{23}
\Omega\equiv\omega_0 +\gamma_2g's\;, \qquad
\Gamma\equiv\gamma_2 (1 -gs) \; ,
\end{equation}
where the effective atomic coupling parameters are
\begin{equation}
\label{24}
g\equiv\;\frac{3\gamma}{4\gamma_2N}\; \sum_{i\neq j}^N \;
\frac{\sin k_0r_{ij}}{k_0r_{ij}}\; , \qquad
g'\equiv\;\frac{3\gamma}{4\gamma_2N}\; \sum_{i\neq j}^N \;
\frac{\cos k_0r_{ij}}{k_0r_{ij}}\; ,
\end{equation}
and the notation
$$
\gamma\equiv\; \frac{4}{3}\; k_0^3d_0^2\; , \qquad
 k_0\equiv \frac{\omega_0}{c} \; , \qquad {\bf d}\equiv d_0{\bf e}_d
$$
is employed. Then from Eqs. (18) to (20) in the mean-field approximation,
we obtain
\begin{equation}
\label{25}
\frac{du}{dt} =\; -\left ( i\Omega +\Gamma\right ) u + s\xi +
\gamma_2(g + ig')su^*{\bf e}_d^2\; ,
\end{equation}
$$
\frac{ds}{dt} =\; - 4\gamma_2 g|u|^2 - 2(u^*\xi +\xi^* u) -\gamma_1
(s-s_0)-
$$
\begin{equation}
\label{26}
- 2\gamma_2\left [ (g+ig')(u^*{\bf e}_d)^2 +
(g-ig')({\bf e}_d^* u)^2 \right ]\; ,
\end{equation}
\begin{equation}
\label{27}
\frac{d|u|^2}{dt} =\; - 2\Gamma |u|^2 +s(u^*\xi +\xi^* u) +
\gamma_2 s \left [ (g+ig')(u^*{\bf e}_d)^2 +
(g-ig')({\bf e}_d^* u)^2 \right ]\; .
\end{equation}

Equations (25) to (27) form a system of nonlinear stochastic differential 
equations. It is worth stressing that the rotating wave approximation 
(RWA), well-known in resonance optics [20,23,25,27], cannot be blindly 
applied for simplifying this system of equations. Really, the standard 
usage of RWA is as follows. One assumes that the polarization variable 
$u$ behaves approximately as $\exp(-i\Omega t)$, that is, rotates with 
the frequency $\Omega$, while the population difference $s$ does not 
rotate. With this assumption, one neglects in the right-hand sides of the 
evolution equations the terms that behave qualitatively differently to 
the assumed behaviour of the left-hand sides. Thus, in Eq. (25) one 
should drop the counter-rotating term containing $u^*$ as well as the
non-rotating term $s\xi$. Similarly, in Eqs. (26) and (27) for 
non-rotating quantities, one should neglect the rotating terms with $u^2$ 
and $(u^*)^2$ as well as with $u\xi^*$ and $\xi^* u$. But, as is evident, 
such a procedure would completely loose information of polariton degrees 
of freedom $\xi$. Hence, RWA is not applicable to such stochastic 
differential equations. Although the RWA reasoning is not justified for 
these equations, one may try to simplify the latter by formally omitting 
only those terms that would be dropped if the polariton field $\xi$ were 
absent, but leaving untouched the terms containing $\xi$. Then one would 
come to the system of equations
$$
\frac{du}{dt} = - (i\Omega +\Gamma)\; u + s\xi \; , 
$$
$$
\frac{ds}{dt} = - 4\gamma_2 g|u|^2 - 2(u^*\xi +\xi^* u) -
\gamma_1( s - s_0) \; ,
$$
$$
\frac{d|u|^2}{dt} = - 2\Gamma|u|^2 + s(u^*\xi +\xi^* u) \; .
$$
This is yet a system of nonlinear, because of Eqs. (23), stochastic 
differential equations.

To solve this system of equations (25) to (27), we shall use the scale 
separation approach [30--32] which is a variant of the Krylov-Bogolubov 
averaging method [33] generalized to the case of stochastic differential 
equations. To this end, we need, first of all, to classify the functional 
variables of the system (25) to (27) onto fast and slow. This can be done 
by accepting the usual inequalities
\begin{equation}
\label{28}
\frac{\gamma_1}{\Omega}\ll 1\; , \qquad \frac{\gamma_2}{\Omega}\ll 1\; ,
\qquad \left | \frac{\Gamma}{\Omega}\right | \ll 1
\end{equation}
and assuming that the average energy of atomic interactions with the
polariton field is also much less than $\Omega$. Then from Eqs. (25) to
(27) it immediately follows that the variable $u$ is to be classified as
fast as compared to the slow variables $s$ and $|u|^2$. The next step is
to solve Eq. (25) for the fast variable treating the slow variables as
quasi-invariants, which yields
\begin{equation}
\label{29}
u(t) =\left [ u_0 + s\int_0^t \; e^{(i\Omega+\Gamma)t'}\; \xi(t')\; dt'
\right ]\; e^{-(i\Omega+\Gamma)t} \; ,
\end{equation}
where $u_0\equiv u(0)$. This solution is to be substituted into the
right-hand sides of the equations (26) and (27) for the slow variables,
with averaging these right-hand sides over polariton degrees of freedom
and over explicitly entering time. In accomplishing this procedure, we
define the effective atom-polariton coupling
\begin{equation}
\label{30}
\alpha\equiv \; \frac{{\rm Re}}{s\Gamma}\; \lim_{\tau\rightarrow\infty}\;
\frac{1}{\tau}\; \int_0^\tau\; \ll u^*(t)\xi(t)\gg\; dt\; ,
\end{equation}
where ${\rm Re}$ means the real part and $\ll\ldots\gg$ implies the
averaging over polariton degrees of freedom. With the form (29) and the
condition $\ll\xi(t)\gg=0$, this gives
$$
\alpha=\; \frac{{\rm Re}}{\Gamma}\; \lim_{\tau\rightarrow\infty}\;
\frac{1}{\tau}\; \int_0^\tau\; dt\; \int_0^t\; e^{-(i\Omega+\Gamma)(t-t')}\;
\ll\xi^*(t)\xi(t')\gg\; dt'\; .
$$
By assumption, the atom-polariton coupling is weak,
\begin{equation}
\label{31}
|\alpha|\ll 1\; .
\end{equation}

In order to understand the structure of the coupling (30), let us model the
polariton field by an ensemble of $N_0$ oscillators, so that
\begin{equation}
\label{32}
\xi(t) =\; \frac{1}{\sqrt{N_0}}\; \sum_k\; \gamma_k\left (
b_ke^{-i\omega_kt} + b_k^\dagger\; e^{i\omega_kt}\right ) \; ,
\end{equation}
where $b_k$ and $b_k^\dagger$ are Bose operators with the properties
$$
\ll b_k^\dagger b_{k'}\gg\; = \delta_{kk'}\; n_k \; , \qquad
\ll b_k b_{k'}\gg\; = 0 \; ,
$$
$$
\ll b_k b_{k'}^\dagger\gg\; =\delta_{kk'}\; (1+n_k) \; , \qquad
\frac{1}{N_0}\; \sum_k\; n_k = 1\; ,
$$
and the effective width $\gamma_k$ is of order $\gamma$ outside the polariton
gap but is zero inside this gap,
$$
\gamma_k=0\; , \qquad \omega_1\;< \omega_k\; <\omega_2\; , \qquad
\Delta_p\equiv  \omega_2-\omega_1\; .
$$
For the polariton field (32), the coupling (30) becomes
\begin{equation}
\label{33}
\alpha=\; \frac{1}{N_0} \sum_k\; |\gamma_k|^2\left [
\frac{n_k}{(\omega_k-\Omega)^2+\Gamma^2} +
\frac{1+n_k}{(\omega_k+\Omega)^2+\Gamma^2}\right ] \; .
\end{equation}
We assume that the collective atomic frequency $\Omega$ lies {\it deeply
inside} the polariton gap of width $\Delta_p$. If this width is much smaller
than the frequencies at the edges of the gap, $\Delta_p\ll\omega_1$, then
Eq. (33) can be approximated by the form
$$
\alpha\approx\; \frac{4\gamma^2}{\Delta_p^2+4\Gamma^2} \; .
$$

After substituting Eq. (29) into the right-hand sides of Eqs. (26) and (27),
with accomplishing the described averaging and introducing the new function
\begin{equation}
\label{34}
w\equiv\; \ll |u|^2\gg - \alpha\; s^2\; ,
\end{equation}
we obtain the equations
\begin{equation}
\label{35}
\frac{ds}{dt} =  - 4g\gamma_2w -\gamma_1(s-s_0)\; , \qquad
\frac{dw}{dt} = -2\gamma_2(1-gs)w
\end{equation}
for the slow variables.

Let us note that the system of three equations (25) to (27) has been derived
from the set of $3N$ equations (18) to (20), where $i=1,2,\ldots,N$, by 
employing a mean-field type approximation for the averages (21). As is 
obvious, it is impossible to deal with a system of $3N$ equations for 
realistically large $N\rightarrow\infty$, because of which one always 
has to invoke some approximation permitting one to reduce the system of 
untreatably large number of equations to a treatable finite dimensional 
dynamical system. The most common such a way of reduction is by using the 
uniform approximation which assumes that the variables $u_i$ and $s_i$ do 
not depend on the index $i$. This approximation is known [20,25] to give
a good description when there are no electromagnetic spatial structures 
and the considered sample is sufficiently large for boundary effects 
being neglected. The uniform approximation reduces Eqs. (18) to (20) to 
the same system of equations (25) to (27). Therefore, the usage of the 
averages (21), together with the mean-field approximation, is mathematically
identical to the uniform approximation. However, from the physical point 
of view, dealing with the averages is more preferable since an averaging 
description serves as a reasonable first approximation even when the 
sample is nonuniform [20,25].

\section{Liberation of Light}

Suppose that at the initial time the atoms are excited so that their
average population difference $s_0\equiv s(0)>0$. For a single atom, when
$g=0$, from Eqs. (35) one gets $s(t)=s_0$, that is, the emission is 
suppressed. For a collective of atoms, when $g\neq 0$, the system of 
equations (35) becomes nonlinear, and its solutions essentially depend on 
the value of the atom-atom coupling parameter $g$, defined in Eq. (24).

According to the notion of suppressed emission, discussed in the
Introduction, we need, first, to consider the stationary solutions of
the evolution equations. For Eqs. (35), there are two stationary solutions.
One of them,
\begin{equation}
\label{36}
s_1^*=s_0 \; , \qquad w_1^*=0 \; ,
\end{equation}
tells that emission is suppressed. And another one,
\begin{equation}
\label{37}
s_2^* =\; \frac{1}{g} \; , \qquad
w_2^*=\; \frac{\gamma_1(gs_0-1)}{4\gamma_2g^2} \; ,
\end{equation}
shows that light can, at least partially, be liberated. The asymptotic
stability of the stationary solutions (36) and (37) can be studied
involving the Lyapunov analysis. This is done by calculating the Jacobian
matrix associated with Eqs. (35) and finding its eigenvalues that are
$$
\lambda^\pm = -\; \frac{1}{2} [\gamma_1 +\gamma_2(1-gs)] \mp\;
\frac{1}{2}\left\{\left [ \gamma_1 -2 \gamma_2(1-gs) \right ]^2 -
32\gamma_2^2 g^2 w\right\}^{1/2} \; .
$$
These eigenvalues, evaluated at the corresponding fixed points (36) and
(37), define the characteristic exponents
$$
\lambda_1^+ = -\gamma_1\; , \qquad \lambda_1^- = -2\gamma_2(1-gs_0) \; ,
$$
\begin{equation}
\label{38}
\lambda_2^\pm =-\; \frac{\gamma_1}{2}\left\{ 1 \pm \left [ 1 +
8\; \frac{\gamma_2}{\gamma_1}\; (1-gs_0)\right ]^{1/2}\right\} \; .
\end{equation}
The real parts of the expressions in Eq. (38) are the Lyapunov exponents
whose signs characterize the stability of the corresponding fixed points.

The Lyapunov analysis shows that if $gs_0<1$, the stationary solution (36)
is a stable node, while the fixed point (37) is a saddle point. When
$gs_0=1$, both fixed points, (36) and (37), merge together becoming neutral.
In the interval
$$
1< gs_0 \leq 1 +\frac{\gamma_1}{8\gamma_2} \; ,
$$ solution (36) is a saddle point, while the fixed point (37) is a stable
node. When
$$
gs_0 > 1 +\frac{\gamma_1}{8\gamma_2}\; ,
$$
the fixed point (36) remains a saddle point, and solution (37) becomes a
stable focus with the characteristic exponents
$$
\lambda^\pm_2 = -\; \frac{\gamma_1}{2} \mp\; i\omega_{osc} \; ,
$$ where the oscillation frequency is
$$
\omega_{osc} \equiv\frac{\gamma_1}{2} \left [ 8\;
\frac{\gamma_2}{\gamma_1}\; (gs_0 - 1) - 1\right ] ^{1/2} \; .
$$
This frequency defines the asymptotic oscillation period
\begin{equation}
\label{39}
T_{osc} \equiv \frac{2\pi}{\omega_{osc}} =
\frac{4\pi}{\sqrt{8(gs_0-1)\gamma_1\gamma_2-\gamma_1^2}}\; .
\end{equation}
In the case when $\gamma_1\ll\gamma_2,\; s_0\sim 1$, and $g\gg 1$, the
latter simplifies to
$$
T_{osc} \simeq 2\pi\sqrt{\frac{T_1T_2}{2g}}\; ,
$$
where $\gamma_1T_1\equiv 1$ and $\gamma_2T_2\equiv 1$, so that $T_1$ and
$T_2$ are the longitudinal and transverse relaxation times, respectively.

In this way, if the coherent interactions between atoms are weak, and the
atom-atom coupling parameter $g$ is small, so that $gs_0<1$, then the system
tends to the stationary solution (36), that is, there is no liberation of
light. When the atomic interactions become stronger, so that $gs_0>1$, the
system tends to another stable stationary solution (37), and a partial
liberation of light occurs since
$$
s_2^* =\frac{1}{g}\; < s_0 \qquad (gs_0 > 1) \; .
$$
The portion of excitation that remains in the atomic ensemble decreases, 
with increasing $g$,
as $s_2^*\rightarrow 0$. The qualitative change of the asymptotic behaviour
of solutions to Eqs. (35) happens when $gs_0=1$. This equality defines the
bifurcation point for Eqs. (35), when the dynamical system is structurally
unstable [34]. This bifurcation point separates the regions where emission
remains suppressed ($gs_0<1$) and where light becomes partially liberated
$(gs_0>1)$. For sufficiently strong atomic interactions, such that
$gs_0>1+\gamma_1/8\gamma_2$, the liberation of light occurs by means of
a series of pulses that, at asymptotically large times, are separated one
from another by the period (39).

To describe the coherent pulse, occurring when $\gamma_1=0$, we
may consider Eqs. (35) omitting there the relaxation term with $\gamma_1$.
Then this system of nonlinear equations can be solved exactly resulting in
the solution
\begin{equation}
\label{40}
s=-\; \frac{\gamma_0}{g\gamma_2}\; {\rm tanh}\left (
\frac{t-t_0}{\tau_0}\right ) +\frac{1}{g} \; , \qquad
w = \frac{\gamma_0^2}{4g^2\gamma_2^2}\; {\rm sech}^2\left (
\frac{t-t_0}{\tau_0}\right )\; ,
\end{equation}
where the radiation width $\gamma_0$ is given by the relation
\begin{equation}
\label{41}
\gamma_0^2 =\Gamma_0^2 + 4g^2\gamma_2^2\left (|u_0|^2 -\alpha_0 s_0^2
\right )
\end{equation}
in which $u_0\equiv u(0)$, $\alpha_0\equiv\alpha(0)$, and
$$
\Gamma_0\equiv \gamma_2(1-gs_0)\; , \qquad \gamma_0 \equiv\frac{1}{\tau_0}\; ,
$$
and where the delay time is
\begin{equation}
\label{42}
t_0 =\frac{\tau_0}{2}\; \ln\left |
\frac{\gamma_0-\Gamma_0}{\gamma_0+\Gamma_0}\right | \; .
\end{equation}
Introducing the critical atom-polariton coupling
\begin{equation}
\label{43}
\alpha_c \equiv\frac{(gs_0-1)^2}{4g^2s_0^2} +\frac{|u_0|^2}{s_0^2}\; ,
\end{equation}
the radiation width can be written as
\begin{equation}
\label{44}
\gamma_0 = 2g|s_0|\gamma_2\sqrt{\alpha_c-\alpha_0} \; .
\end{equation}
The value (43) is termed critical since the coupling $\alpha_0$ cannot
exceed $\alpha_c$ for the solution (40) to remain finite. The restriction
$\alpha_0\leq\alpha_c$ specifies condition (31) assumed for the validity of
the averaging method. This restriction is not as severe as far as
$\alpha\ll 1$ while, for large $g$, $\alpha_c\geq 1/4$.

At the time $t=t_0$, the solutions (40) and (34) yield
\begin{equation}
\label{45}
s(t_0) =\frac{1}{g} \; , \qquad w(t_0) = s_0^2(\alpha_c-\alpha_0)\; ,
\qquad
\ll |u(t_0)|^2\gg\; = s_0^2(\alpha_c -\alpha_0) +\frac{\alpha}{g^2} \; .
\end{equation}
Then the system of atoms achieves the maximum of coherence.

Consider more in detail the case when the system of atoms is initially
completely inverted, $s_0=1$, there is no triggering pulse, $u_0=0$, and
the atom-atom coupling is strong, $g>1$. Then the critical value (43)
becomes
\begin{equation}
\label{46}
\alpha_c =\left (\frac{g-1}{2g}\right )^2 \; .
\end{equation}
If $\alpha\ll\alpha_c$, the radiation width (44) gives
\begin{equation}
\label{47}
\gamma_0 = (g-1)\gamma_2\left [ 1 -\;\frac{2g^2\alpha_0}{(g-1)^2}\right ]\; .
\end{equation}
The corresponding radiation time is
\begin{equation}
\label{48}
\tau_0 =\frac{T_2}{g-1} \left [ 1 +\frac{2g^2\alpha_0}{(g-1)^2}\right ]\; .
\end{equation}
And for the delay time (42), we have
\begin{equation}
\label{49}
t_0 =\frac{T_2}{2(g-1)}\; \ln\left |
\frac{(g-1)^2-g^2\alpha_0}{g^2\alpha_0}\right | \; .
\end{equation}
When the atomic coupling is strong, then
\begin{equation}
\label{50}
\alpha_c\simeq \frac{1}{4} \qquad (u_0=0,\; g\gg 1) \; .
\end{equation}
The radiation time (48) for $g\gg 1$ becomes
\begin{equation}
\label{51}
\tau_0 \simeq \frac{T_2}{g}\; (1+2\alpha_0) \; .
\end{equation}
Hence, $\tau_0$ can be much smaller than $T_2$. Since $g\sim N$, we have
$\tau_0\sim N^{-1}$, which is typical of superradiance [20,25]. Under
condition (50), the delay time (49) is
\begin{equation}
\label{52}
t_0 \simeq \frac{T_2}{2g} \; |\ln\alpha_0| \; .
\end{equation}
Note that if $\alpha_0\rightarrow 0$, then $t_0\rightarrow\infty$, which
means that emission would be suppressed for very long time. The 
superradiant burst develops at the delay time $t_0$ and lasts during the
radiation time $\tau_0$.

In the case when $\alpha_c-\alpha_0\ll 1$, the radiation time increases, as
compared to the opposite case when $\alpha_0\ll\alpha_c$, being
\begin{equation}
\label{53}
\tau_0 = \frac{T_2}{2g|s_0|\sqrt{\alpha_c-\alpha_0}} \; ,
\end{equation}
while the delay time shortens becoming
\begin{equation}
\label{54}
t_0 =\frac{2g|s_0|T_2}{(gs_0-1)^2} \; \sqrt{\alpha_c-\alpha_0}\; .
\end{equation}
In this case, the radiation can be coherent if $\tau_0\ll T_2$, which
requires $g\sqrt{\alpha_c-\alpha_0}\gg 1$.

The superradiant character of emission is connected with the radiation time
being inversely proportional to the number of radiators, $\tau_0\sim N^{-1}$.
Another characteristic describing the level of coherence in the radiating
system is the radiation intensity. The total intensity of radiation,
averaged over fast oscillations,
\begin{equation}
\label{55}
I(t) = I_{inc}(t) + I_{coh}(t) \; ,
\end{equation}
consists of two terms, the intensity of incoherent radiation
\begin{equation}
\label{56}
I_{inc} =\frac{1}{2}\;\omega_0\gamma N(1+s)
\end{equation}
and the intensity of coherent radiation
\begin{equation}
\label{57}
I_{coh} =\omega_0\gamma\varphi_s N^2 \ll|u|^2\gg \; ,
\end{equation}
where $\varphi_s$ is the shape factor [35] given by the integral over
spherical angles,
\begin{equation}
\label{58}
\varphi_s=\frac{3}{8\pi} \; \int\; \left |{\bf n}\times{\bf e}_d\right |^2\;
F(k_0{\bf n})\; d\Omega({\bf n}) \; , \qquad
F({\bf k}) \equiv \left | \frac{1}{N}\;
\sum_{i=1}^N \; e^{i{\bf k}\cdot{\bf r}_i}\right |^2 \; .
\end{equation}
For the point-like system, for which $k_0\rightarrow 0$, we have $\varphi_s
\rightarrow 1$, and the intensity of coherent radiation is proportional to
the number of radiators squared, $I_{coh}\sim N^2$, which is typical of
the Dicke model. For a finite-size system, the shape factor essentially
depends on the relation between the radiation wavelength $\lambda$ and the
characteristic sizes of the sample [35]. Thus, for a cylindrical sample of
radius $R$ and length $L$, we have
\begin{eqnarray}
\varphi_s\simeq \left\{ \begin{array}{ccc}
\frac{3\lambda}{8L}\; , &\frac{\lambda}{2\pi L}\ll 1\; ,
& \frac{R}{L}\ll 1\; , \\
\nonumber
\frac{3}{8}\left (\frac{\lambda}{\pi R}\right )^2\; ,
& \frac{\lambda}{2\pi R}\ll 1\; , & \frac{L}{R}\ll 1\;
\end{array} \right.
\end{eqnarray}
for pencil-like or disk-like shapes, respectively. Hence for the first case
$I_{coh}\sim N^{5/3}$, while for the second case $I_{coh}\sim N^{4/3}$.

To compare the intensities of coherent and incoherent radiation, it is
convenient to introduce the coherence coefficient [36] defined as the ratio
\begin{equation}
\label{59}
C_{coh}(t) \equiv \frac{I_{coh}(t)}{I_{inc}(t)} \; .
\end{equation}
For the coherent radiation intensity (57), using the relation (34), we get
$$
I_{coh}(t) =\omega_0\gamma\varphi_s N^2(w+\alpha s^2) \; .
$$
Therefore, the coherence coefficient (59) is
\begin{equation}
\label{60}
C_{coh} = 2\varphi_s N \; \frac{w+\alpha s^2}{1+s}\; .
\end{equation}
At the moment of the maximal coherence of the superradiant burst,
according to Eqs. (45), we find
$$
C_{coh}(t_0) = 2g\varphi_s s_0^2 N \; \frac{\alpha_c-\alpha_0}{1+g} \; .
$$
If $C_{coh}>1$, the radiation is predominantly coherent. For the case
when $u_0=0,\; s_0=1$, and $g\gg 1$, the latter expression gives
$$
C_{coh}(t_0)\simeq \frac{1}{2}\; \varphi_s N \; .
$$
If the sample has a pencil-like or disk-lake shape, then
\begin{eqnarray}
C_{coh}(t_0) \simeq \left\{ \begin{array}{cc}
\frac{3\pi}{16}\left (\frac{R}{\lambda}\right )^2\; \rho\lambda^3\; ,
& pencil \\
\nonumber
\frac{3}{16\pi}\left (\frac{L}{\lambda}\right ) \; \rho\lambda^3\; ,
& disk\; , \end{array}\right.
\end{eqnarray}
where $\rho$ is the density of resonance atoms. When this density is
sufficiently high, and $\lambda\ll R$ or $\lambda\ll L$, the coherence
coefficient can be very large, which would mean that the radiation is almost
purely coherent. 

The general condition for light to be, at least partially, liberated is
\begin{equation}
\label{61}
s(\infty) < s_0 \; .
\end{equation}
From equations (40), we have
$$
s(\infty) = -\; \frac{\gamma_0}{g\gamma_2} \; + \; \frac{1}{g} \; ,
\qquad w(\infty) = 0 \; ,
$$
which, with the radiation width (44), gives
\begin{equation}
\label{62}
s(\infty) = -2|s_0|\sqrt{\alpha_c-\alpha_0} \; + \; \frac{1}{g} \; .
\end{equation}
Therefore the liberation condition (61) becomes
\begin{equation}
\label{63}
g(s_0 + 2|s_0|\sqrt{\alpha_c-\alpha_0}) > 1 \; .
\end{equation}
In the case, when $u_0=0$, and taking into account that $\alpha_0\ll 1$, 
equation (62) reduces to
\begin{eqnarray}
s(\infty) =\left\{ \begin{array}{cc}
s_0\; , & gs_0 < 1 \\
-s_0 + 2/g \; , & gs_0 > 1 \; . \end{array} \right.
\end{eqnarray}
The condition (63) simplifies to
\begin{equation}
\label{65}
gs_0 > 1 \; .
\end{equation}
It is interesting that, although the limit values $s(\infty)$ are 
different for the cases when $\gamma_1$ is finite or when it is zero, but 
the liberation condition (65) remains the same. Under this condition, a 
system of atoms can radiate, though spontaneous emission of a single atom 
is suppressed. The case of a single atom can be recovered by setting 
$g\rightarrow 0$. Then, as is obvious, condition (65) can never hold 
true. It is only when the density of doped atoms is sufficiently high, so 
that the atom-atom coupling $g$ becomes sufficiently large, satisfying 
condition (65), the radiation of atoms is possible being due to 
collective effects.

\section{Discussion}

We considered a system of resonance atoms doped into a medium with a
polariton band gap. Spontaneous emission of a single atom with the 
transition frequency inside the polariton gap is suppressed.
However an ensemble of atoms with their transition frequencies in
the gap can radiate due to effective coherent interactions between the 
atoms. If this intertaction is sufficiently strong, light is partially 
liberated. The collective liberation of light occurs through one or a 
series of superradiant pulses.

The dynamics of light liberation has been analysed employing the scale
separation approach [30--32], which is a generalization of the averaging
method [33] to stochastic differential equations. This approach provides
an efficient tool for treating complicated systems of nonlinear evolution
equations, as has earlier been demonstrated for the problems of superradiant
spin relaxation in nonequilibrium magnets [30--32] and of nonlinear dynamics
of atoms in magnetic traps [37--39], where nonlinear phenomena are of
crucial importance. Collective liberation of light is also a principally
nonlinear phenomenon, with nonlinearity caused by coherent atomic
interactions.

In conclusion, let us give some estimates for the characteristic parameters
considered in the text. The polariton effect is well developed in many
dielectrics and semiconductors [4,5]. For instance, it is intensively
studied in such semiconductors as CuCl, CuBr, CdSe, ZnSe, GaAs, GaSb,
InAs, AlAs, and SiC [10--12,40]. The polariton gap in such materials 
develops around the frequency $10^{14}$ s$^{-1}$, with the gap width
$\Delta_p\sim 10^{13}$ s$^{-1}$. By assumption, the atomic transition
frequency is inside the polariton gap, i.e. $\omega_0\sim 10^{14}$ s$^{-1}$.
Hence the radiation wavelength is $\lambda\sim 10^{-3}$ cm. For the 
line width, one may take $\gamma_2\sim 10^9$ s$^{-1}$.

For the initially inverted atoms, with $s_0=1$, the critical atom-atom
coupling parameter, above which the collective radiation becomes possible,
is $g_c=1$. For the parameters $g>g_c$, the collective liberation of light
occurs. If $\gamma\sim\gamma_2$, then the atomic coupling parameters,
defined in Eq. (24), are $g\sim\rho\lambda^3$ and $g'\sim\rho\lambda^3$,
where $\rho$ is the density of atoms. Consequently, the critical density
of atoms, providing $g_c=1$, is $\rho_c\sim\lambda^{-3}$, which gives
$\rho_c\sim 10^9$ cm$^{-3}$. When $\rho>\rho_c$, radiation becomes possible
because of the formation of an impurity band inside the polariton band. The
width of the impurity band is of the order of the collective line width
$\Gamma\sim\gamma_2g\sim\gamma\rho\lambda^3$. The latter becomes larger
than the polariton gap, when the density $\rho>\Delta_p/\gamma_2\lambda^3$,
that is, $\rho> 10^{13}$ cm$^{-3}$. For such densities of resonance atoms,
the polariton gap can be overlapped by the impurity band. At the density 
$\rho\sim 10^{13}$ cm$^{-3}$, the atom-atom coupling is $g\sim 10^4$, and 
the collective line width is $\Gamma\sim 10^{13}$ s$^{-1}$.

With the densities $\rho<10^{13}$ cm$^{-3}$, the atom-atom coupling $g<10^4$,
while the atom-polariton coupling is $\alpha\sim\gamma^2/\Delta_p^2$, that
is $\alpha\sim 10^{-8}$. The radiation time $\tau_0\sim T_2/g$ is larger than
$10^{-13}$ s but can be much smaller than $T_2\sim 10^{-9}$ s$^{-1}$. And
the delay time $t_0\sim\tau_0|\ln\alpha|$ is an order longer than the
radiation time. In this way, for the density of doped atoms $\rho\sim 
10^9-10^{13}$ cm$^{-3}$, the effective atom-atom coupling is $g\sim 1-10^4$.
Then the delay time is $t_0\sim 10^{-12}-10^{-8}$ s and the radiation time 
$\tau_0\sim 10^{-13}-10^{-9}$ s. 

\vskip 5mm

I am grateful to M.R. Singh for useful discussions and hospitality during
my stay at the Centre for Interdisciplinary Studies in Chemical Physics at
the University of Western Ontario, London, Canada, where this paper was
started. I appreciate the Senior Fellowship of the University of Western
Ontario. This work has also been supported by the Bogolubov-Infeld Grant
of the State Agency for Atomic Energy, Poland, and by a Grant of the S\~ao
Paulo State Research Foundation, Brazil.

\end{document}